\documentstyle[aps,multicol,epsf,epsfig]{revtex} 
\newcommand{\xsize}{\epsfxsize=18.0cm}

\begin{document}

%\draft

\title{Fractal Behavior of the Shortest Path Between Two Lines in
Percolation Systems}

\author{Gerald Paul,$^\ast$ Shlomo Havlin,$^\dagger$ and H. Eugene Stanley$^\ast$} 

\address{$^\ast$Center for Polymer Studies and Department of Physics\\
Boston University, Boston, MA 02215 USA\\
$^\dagger$Minerva Center and Department of Physics\\
Bar-Ilan University, Ramat Gan, Israel}

\date{phs.tex ~~~ 04 March 2002 ~~~ draft}

\maketitle

\begin{abstract}
%\abstracts{
Using Monte-Carlo simulations, we determine the scaling form for the
probability distribution of the shortest path, $\ell$, between two lines
in a 3-dimensional percolation system at criticality; the two lines can
have arbitrary positions, orientations and lengths. We find that the
probability distributions can exhibit up to four distinct power law
regimes (separated by cross-over regimes) with exponents depending on
the relative orientations of the lines. We explain this rich fractal
behavior with scaling arguments.
%}
\end{abstract}

\begin{multicols}{2}  
\section{Introduction}

There has been considerable recent activity
\cite{Dokh,Lee60,Lee62,Grassberger,Ziff99} analyzing $P(\ell|r)$, the
probability distribution for the length of shortest path, $\ell$,
between two points separated by Euclidean distance $r$ in a percolation
system \cite{Neumann,Stauffer,BundeHavlin,Ben-Avraham00}. This paper extends that work
by determining the scaling form of the distribution of shortest paths
between two lines of arbitrary position, relative orientation and
lengths in 3-dimensional systems. These configurations are important
because they much more accurately model the configurations used in oil
recovery in which fluid is injected in one well (one of the lines in our
configuration) and oil is recovered at a second well (the second line in
our configuration); the wells may, in reality, be at arbitrary orientation
and of different lengths.

The scaling form for the 2-points configuration in which the 2 points
are located at $((L-r)/2,L/2,L/2),((L+r)/2,L/2,L/2)$ in a system of side
$L$ has been found to be \cite{Dokh}
\begin{equation}
\label{e.1}
P(\ell|r)\sim {1\over  r^{d_{\mbox{\scriptsize
min}}}}    \left( {\ell\over r^{d_{\mbox{\scriptsize
min}}}}\right)^{-g_{\ell}}f_1\left({\ell\over r^{d_{\mbox{\scriptsize
min}}}}\right)f_2\left({\ell\over L^{d_{\mbox{\scriptsize
min}}}}\right),
\end{equation}
where
%
%\begin{mathletters}
\begin{equation}
\label{e.2a}
f_1(x)=e^{-ax^\phi}
\end{equation}
and
\begin{equation}
\label{e.2b}
f_2(x)=e^{-bx^\psi}.
\end{equation}
%\end{mathletters}
%
The exponents $g_\ell$, $d_{\mbox{\scriptsize min}}$, $\phi$, and $\psi$
are universal and the constants $a$ and $b$ depend on lattice type. In
3D, the values of these exponents have previously been found to be
\cite{Lee62} $g_\ell=2.3\pm 0.1$, $d_{\mbox{\scriptsize min}}=1.39\pm
0.05$, $\phi=2.1\pm 0.5$, and $\psi=2.5\pm 0.5$. The first stretched
exponential function, $f_1$, reflects the fact that the shortest path
must always be at least equal to the distance $r$ between the two
points; the second stretched exponential function, $f_2$, reflects the
fact that the lengths of the shortest paths are bounded because of the
finite size, $L$, of the system.

We find that the scaling form for the 2-lines configuration has the same
form as that found for the shortest path distribution between two points
with the exceptions that: (i) the power law regime of the distribution
as represented by the term $(\ell/r^{d_{\mbox{\scriptsize
min}}})^{-g_{\ell}}$ in Eq.~(\ref{e.1}) is replaced with up to four
different power law regimes (separated by cross-over regimes) with
exponents depending on the relative orientations of the lines and (ii)
the Euclidean distance, $r$, in Eq.~(\ref{e.1}) between the two points
is replaced by the shortest Euclidean distance between the two lines.
The lengths of the lines affect the sizes of the power law regimes.

\section{Simulations}

Monte Carlo simulations were performed using the Leath method and
growing clusters from 2 sets of seeds---one for each line. The length of
the shortest path between the two lines is the sum of the chemical
distances from each set of seed sites to the point where a cluster
started at one set of seeds meets a cluster started from the other set
of seeds \cite{Grassberger}.  The cluster growth for a given realization
is terminated when the two clusters meet.  For parallel line
configurations, in which the probability distributions decay rapidly, we
use the method of Ref.\cite{Grassberger02} to obtain good statistics for
shortest paths that have very low probabilities.  We use the memory management
technique described in \cite{Paul01} to perform simulations in which the
growing clusters never hit a boundary of the system.

The clusters that are created and included in our analysis are of all
sizes, not just the incipient infinite cluster.

\section{Non-Parallel Wells}
\subsection{Co-Planar Lines}

\subsubsection{Equal Length Symmetric Lines}

\begin{figure}

\centerline{
\epsfxsize=7.0cm
\epsfclipon
\epsfbox{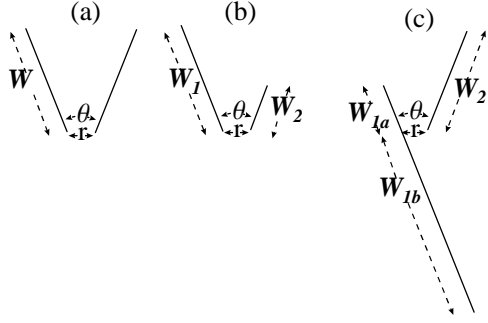}
}

\caption{Example configurations of two non-parallel lines which are
studied. (a) Simple configuration of lines of equal length. (b)
Configuration of lines of unequal length ($W_1>W_2$). (c) Configuration
in which shortest distance between lines does not terminate at the ends of
lines ($W_{a1}<W_2<W_{1b}$).}
\label{pPP1}
\end{figure}

We start by considering relatively simple configurations of the type
shown in Fig.~\ref{pPP1}(a) in which the lines are co-planar, of equal length and
are positioned symmetrically. We will study configurations in which the
lines are of unequal length [see Fig.~\ref{pPP1}(b)] and/or are not positioned
symmetrically [see Fig.~\ref{pPP1}(c)] in the following sections. In all of these
configurations, $r$ is the shortest Euclidean distance between the two
lines.

%\subsubsection{Power Law Regime}

\begin{figure}

\centerline{
\xsize
\epsfclipon
%\epsfbox{pEq1.eps}
}

\centerline{
\xsize
\epsfclipon
\epsfbox{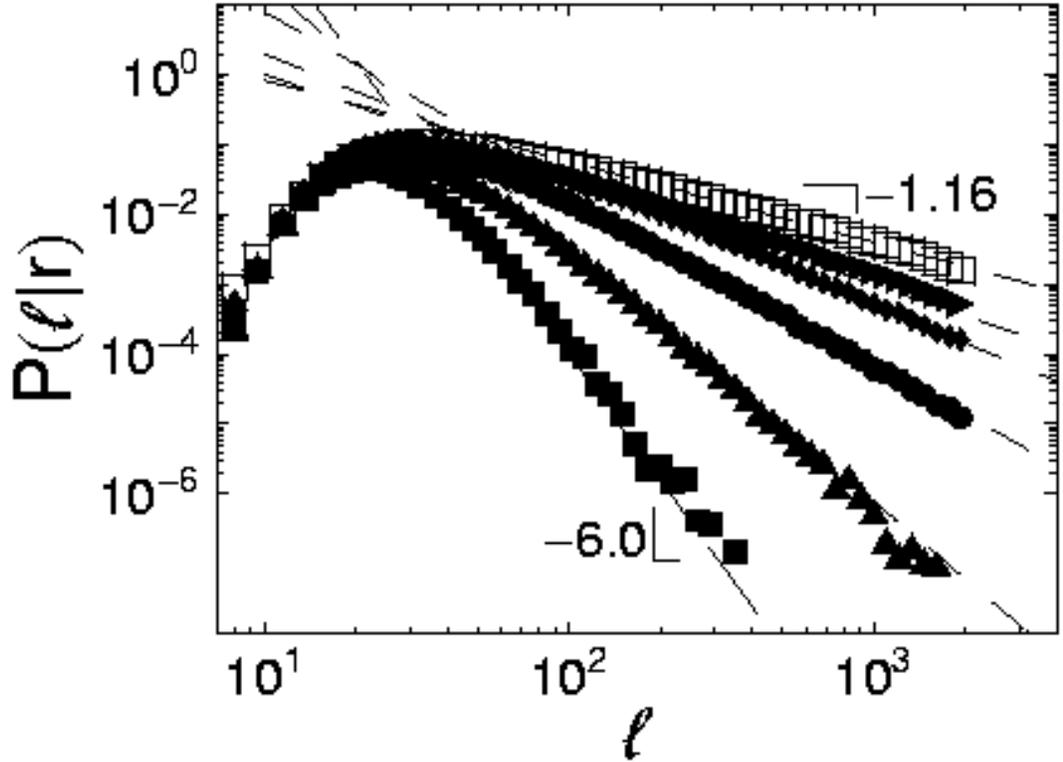}
}
\caption{$P(\ell | r)$ vs $\ell$ for configuration of two lines of equal
length with $r=8$, $\theta=$ (from bottom to top) $3^\circ$ (filled
square), 6$^\circ$, 12$^\circ$, 20$^\circ$, 40$^\circ$, and 180$^\circ$
(unfilled square). The corresponding well lengths W are 4890, 2445,
1224, 737, 374 and 128, respectively. The plots are normalized such that
the initial sections of plots are coincident.}
\label{pEq}
\end{figure}

Figure \ref{pEq} contains log-log plots of $P(\ell|r)$, the
shortest path distribution for $r=8$ and various values of $\theta$. We
have chosen $r=8$ so that the initial cutoff is present; for smaller
values of r, lattice effects destroy this initial cutoff. Since the
focus of this paper is the power law regimes, not the initial or final
cutoffs, in all later figures we will choose configurations with $r=1$
so that the extent of the power law regimes is as long as possible. The
exception to this will be cases in which $\theta$ is very small where
small $r$ introduces other lattice effects.

We note that after the initial peak in each distribution, there is a
power-law regime, the slope of which, $g_\ell(\theta)$, increases with
increasing $\theta$. We will call this power law regime the ``2-lines
regime.'' Simple scaling arguments imply that if the lengths of the
lines were infinite, these 2-lines regimes would continue
indefinitely. For finite line lengths, we would expect that, for large
$\ell$, the distributions would exhibit a crossover to a power law
regime with the same exponent as that for a configuration with two
points---because for large $\ell$ the long paths travel far enough away
from the lines that they appear to be points. In this regime, the power
law exponent has the value of that of two points, 2.35 \cite{Lee62}. For the plots in
Fig.~\ref{pEq}, in order to see the power regimes clearly, we have chosen the
lengths of the lines long enough that this crossover occurs after the
maximum value of $\ell$ in the plots.

\begin{figure}[h]

\centerline{
\xsize
\epsfclipon
\epsfbox{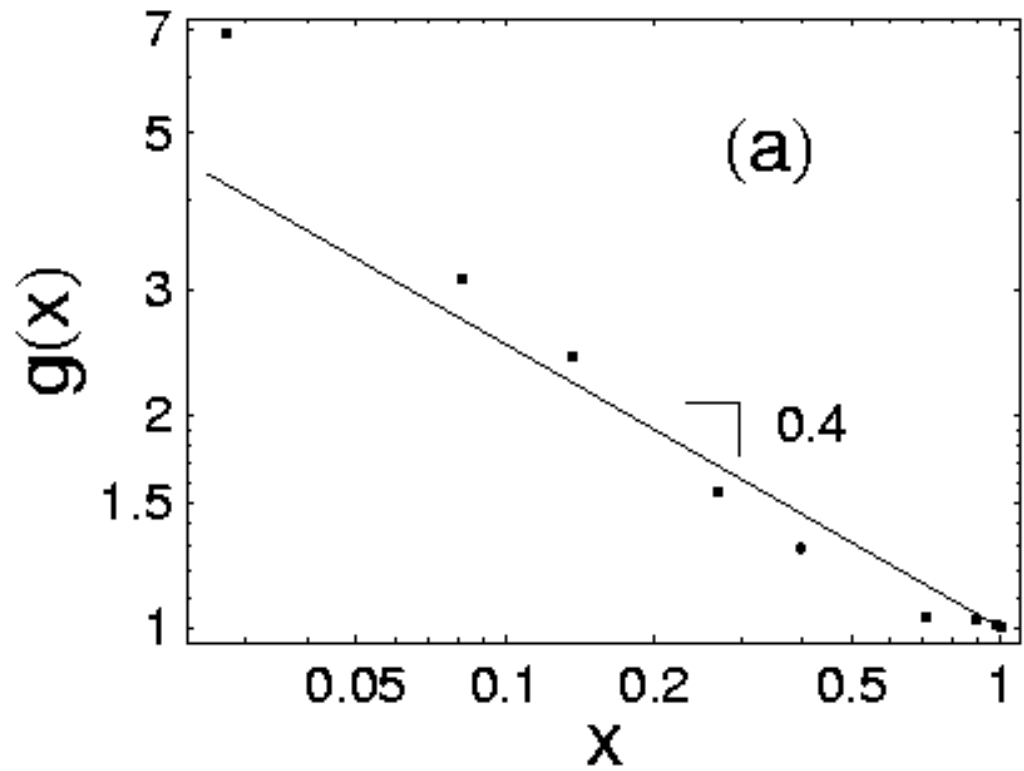}
}

\centerline{
\xsize
\epsfclipon
\epsfbox{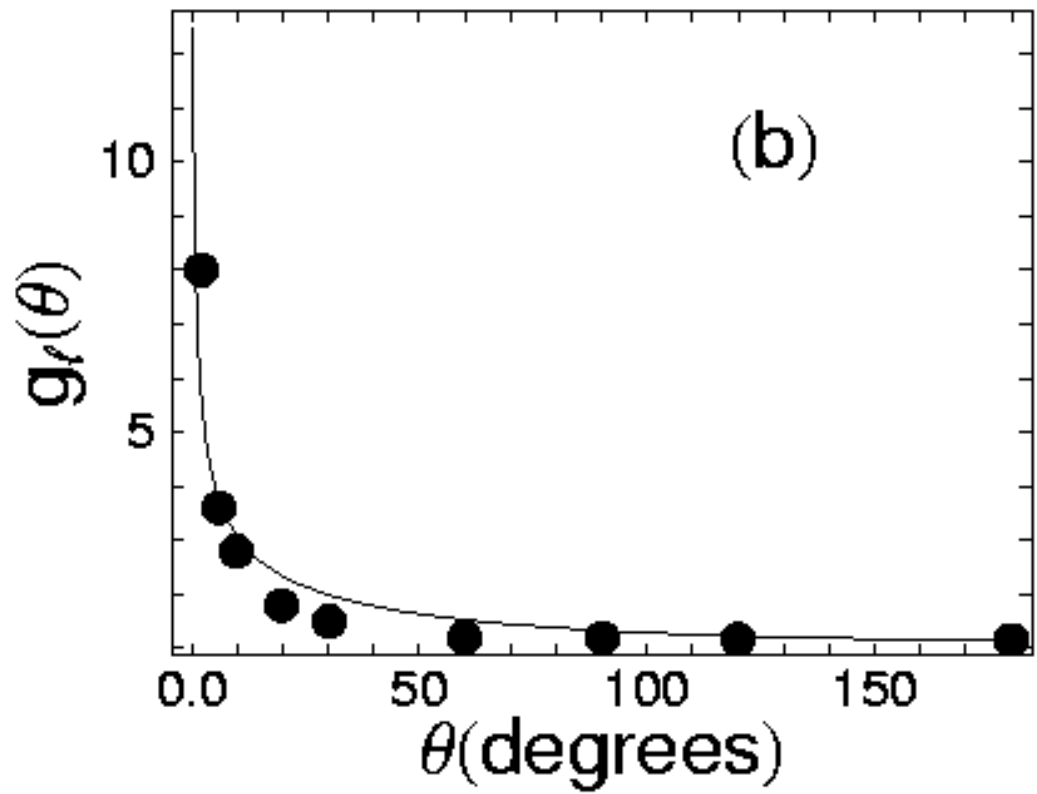}
}

\caption{$g_{\ell}(\theta)$ vs $\theta$. The solid line is a plot of
Eq.~(\protect\ref{e1x}).} 
\label{gl}
\end{figure}

In Fig.~\ref{gl}, we plot $g_\ell(\theta)$ versus $\theta$. The plot
suggests that $g_\ell(\theta)$ diverges as $\theta$ goes to
zero. We attempt to fit this function with a power of $1/sin((\theta)/2)$
and find the best fit with the function
\begin{equation}
\label{e1x}
f(\theta)={g_l(180^\circ)\over\sin(\theta/2)^{0.4}}.
\end{equation}
This form is not based on any fundamental theory; it simply has the
properties that $f(\theta=180^\circ)=g_{\ell}(\theta=180^\circ)$, it
diverges as $\theta$ goes to zero and it fits the intermediate points
reasonably well.

\begin{figure}

\centerline{
\xsize
\epsfclipon
\epsfbox{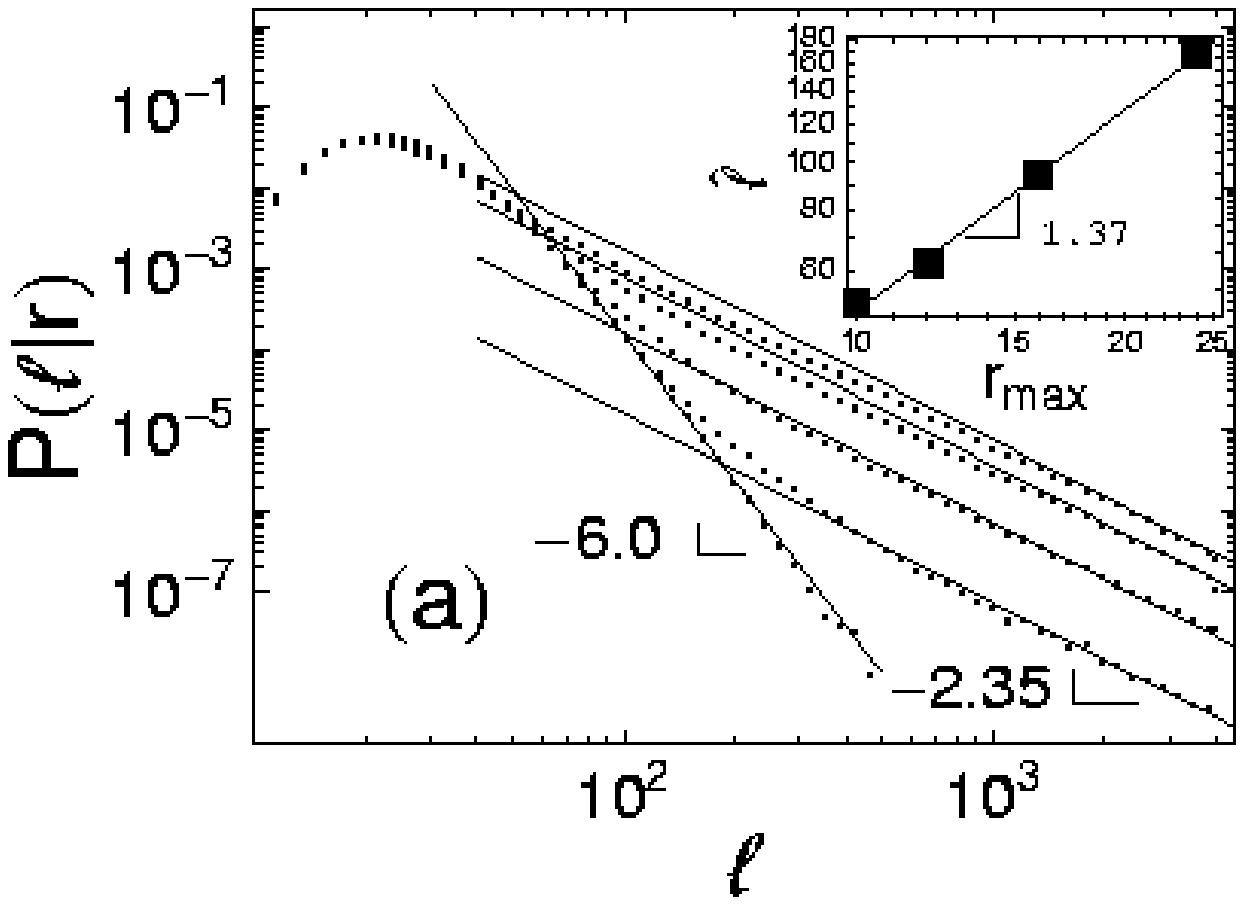}
}

\centerline{
\xsize
\epsfclipon
\epsfbox{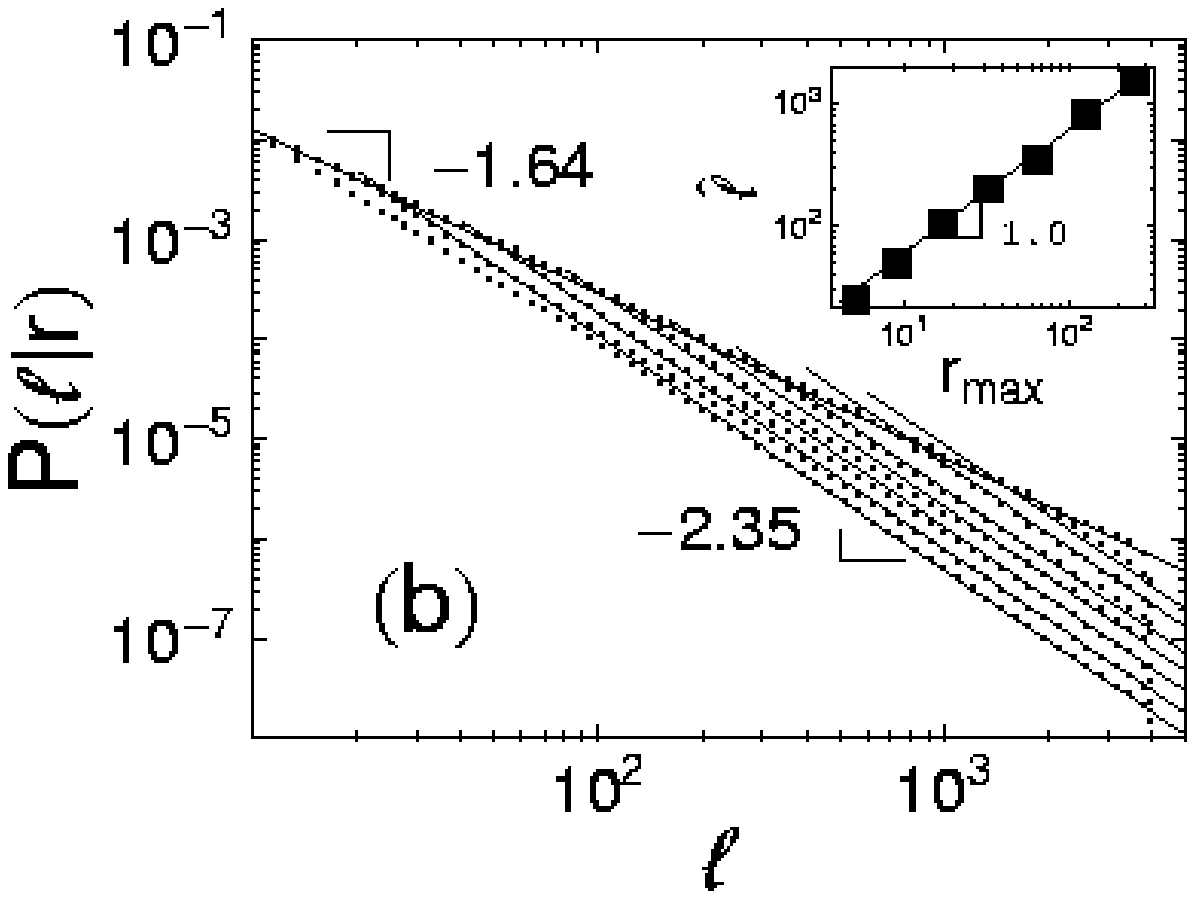}
}

\centerline{
\xsize
\epsfclipon
\epsfbox{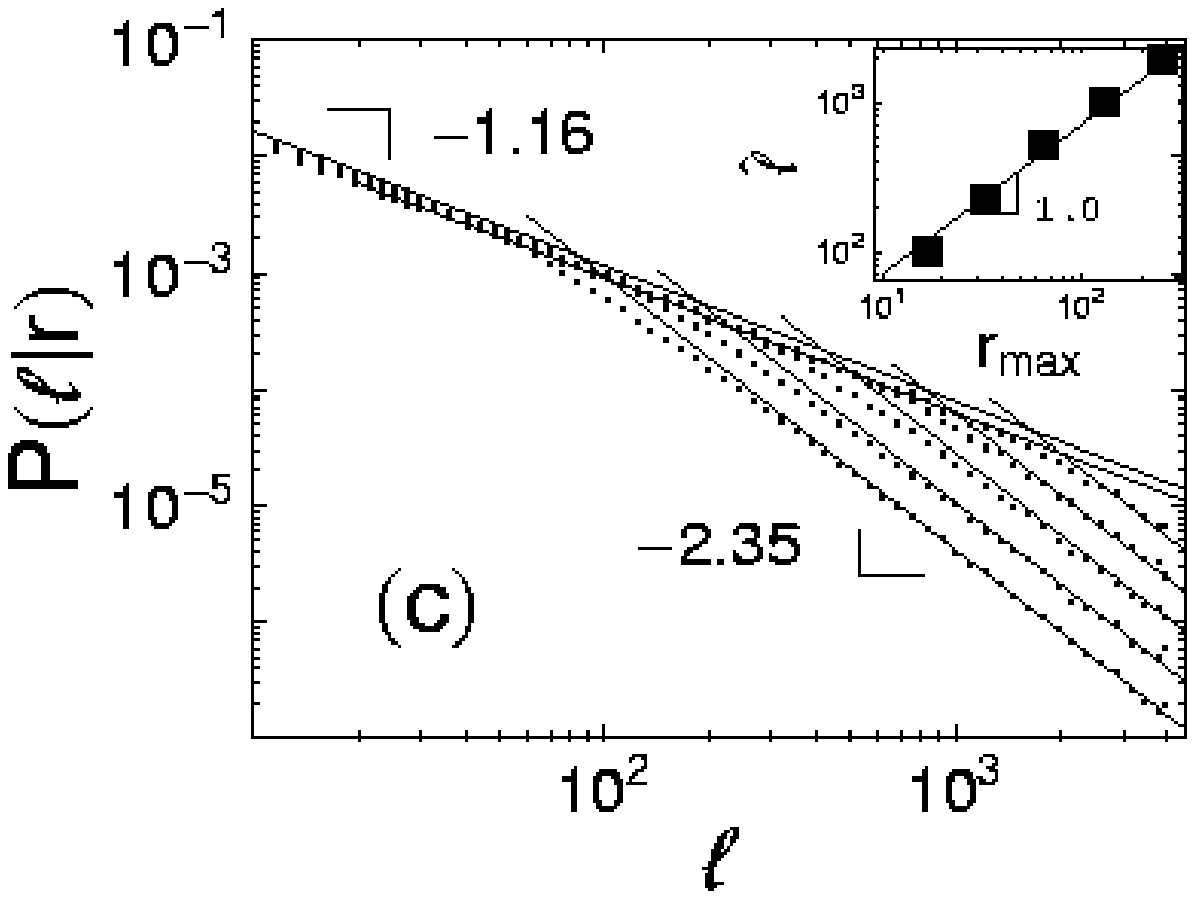}
}
\caption{$P(\ell | r)$ vs $\ell$ for configuration of two lines of equal
length. (a) $r=8$, $\theta=3^\circ$, $W=$ (from top to bottom) 38, 76,
152, 304, 1216, and 2432 (b) $r=1$, $\theta=29^\circ$, $W=$ (from bottom
to top) 8, 17, 33, 66, 132, 265, and 529, (c) $r=1$, $\theta=180^\circ$,
$W=$ (from bottom to top) 8, 16, 32, 64, and 128. For all plots, the
larger the value of $W$, the larger the value of $\ell$ at which
behavior changes from 2-lines behavior to 2-points behavior for which
the slope is $-2.35$. The insets plot the crossover value, $\hat\ell$,
vs. $r_{\mbox{\scriptsize max}}$.}
\label{pEqCutoff}
\end{figure}

The crossover between the 2-line regime and the 2-point regime is
illustrated in Fig.~\ref{pEqCutoff} where in each panel we plot $P(\ell|r)$ for
fixed $\theta$, and various values of $W$. As expected, the larger the
length of the lines, the higher the value of $\hat\ell$, the value of
$\ell$ at which the crossover occurs.  Quantitative analysis of the
crossover behavior is given in Section III.C.

\subsubsection{Point-Line Configurations}

\begin{figure}

\centerline{
\epsfxsize=7.0cm
\epsfclipon
\epsfbox{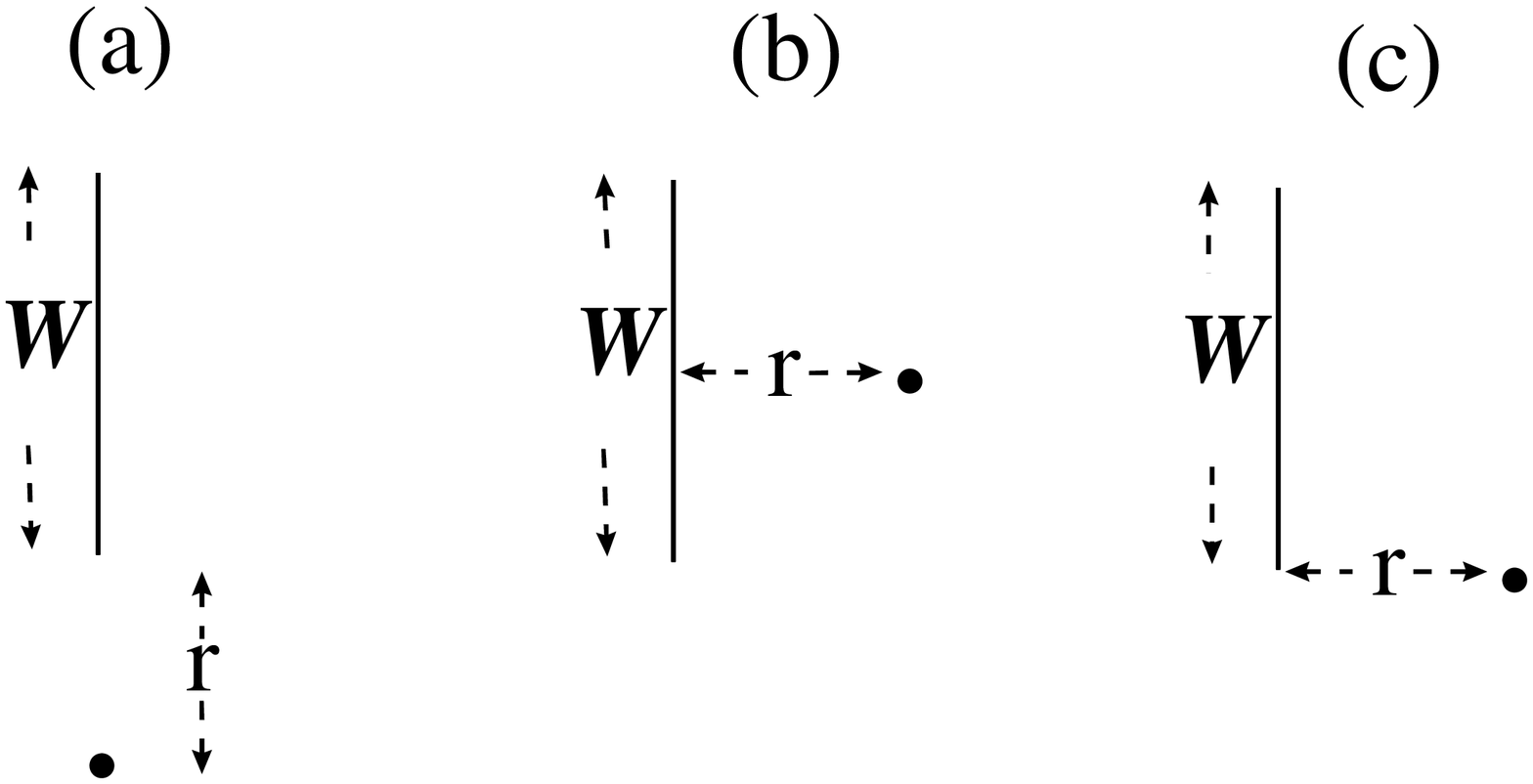}
}
\caption{Example configurations in which one line is of finite length $W$
and one is of zero length (i.e., a point). In all cases, the shortest
distance between the point and the line is $r$.}
\label{pPtLineConfig}
\end{figure}

\begin{figure}
\centerline{
\xsize
\epsfclipon
\epsfbox{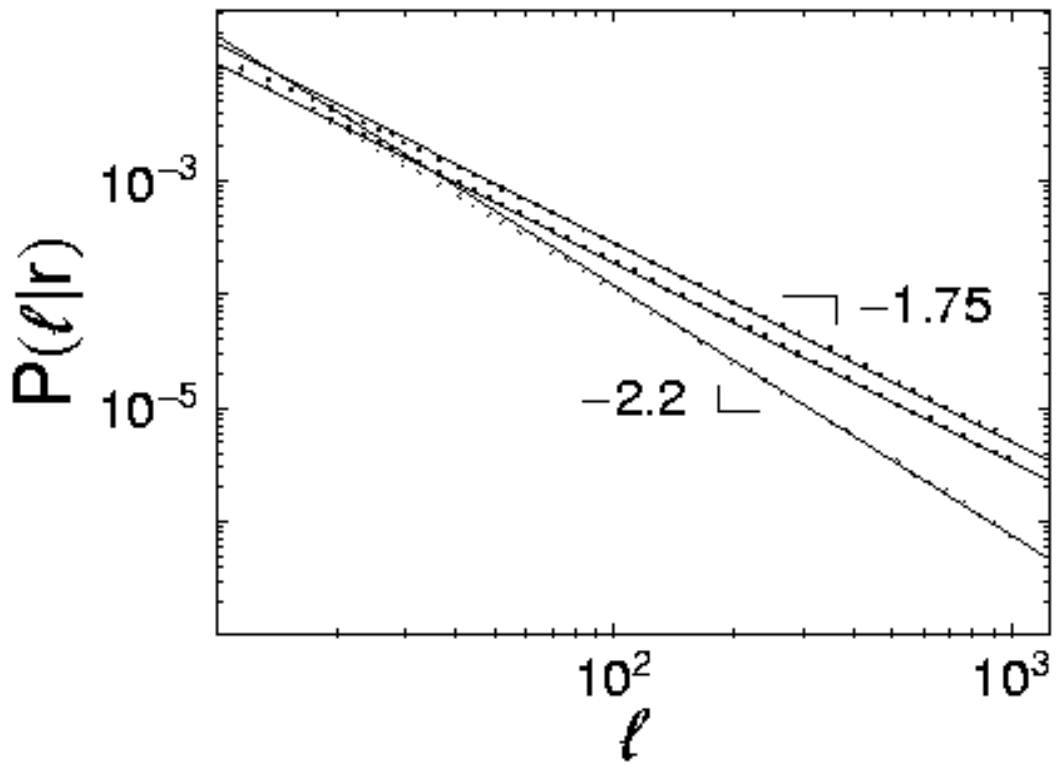}
}
\caption{$P(\ell | r)$ vs $\ell$ for configuration of a point and a line
with $r=1$ and $W=128$. From top to bottom, the plots are for the
configurations shown in Figs.~\protect\ref{pPtLineConfig}(a), (c), and
(b) respectively. We see that the slopes in configurations where the
point is closest to the end of the line
[Fig.~\protect\ref{pPtLineConfig} (a) and (c)] are the same (with some
initial difference) and they are different from the slope in the
configurations in which the point is closest to the middle of the line
(Fig.~\protect\ref{pPtLineConfig}b).}
\label{pPtLine1}
\end{figure}

We next study configurations in which one line has zero length (i.e., a
point) and the other is a line of finite length. This is the extreme
case of the configuration in which the two lines have different
length. We will study the case where both lines have finite length in
the next section. We in fact study the three configurations shown in
Figs.~\ref{pPtLineConfig}(a)--(c). The plots of $P(\ell|r)$ for these
configurations are shown in Fig.~\ref{pPtLine1}. The plots have a
power-law regime with exponent $-1.75$ for the configurations of
Fig.~\ref{pPtLineConfig}(a) and Fig.~\ref{pPtLineConfig}(c), and
exponent $-2.2$ for the configuration of Fig~\ref{pPtLineConfig}(b). We
denote this regime the ``point-line regime.'' Fig.~\ref{pPtLine2} shows
the crossover from point-line behavior to 2-points behavior.

\subsubsection{Unequal Length Symmetric Lines}

We can now study configurations of the type shown in Fig.~1(b) in which
the lines are of different lengths, $W_1$ and $W_2$. For such a
configuration we would expect three power-law regimes: (i) for small
$\ell$ such that the two lines appear to be infinite, a 2-line regime,
with slope dependent on $\theta$, (ii) a point-line regime, with slope
$-1.75$, for values of $\ell$ large enough that the shorter line appears
to be a point, and (iii) a 2-point regime, for even larger values of
$\ell$ where both lines appear to be points. Plots of $P(\ell|r)$ for such
configurations are shown in Fig.~\ref{pDiff} and are consistent with our
expectations.

\subsubsection{Complex Configurations (Unequal Length Non-Symmetric Lines)}

The last of the co-planar configurations is of the type shown in
Fig.~\ref{pPP1}(c). In general, based on the reasoning above, for configurations
of this type we would expect $P(\ell|r)$ to have four power-law
regimes. For the configuration shown in Fig.~\ref{pPP1}(c), in which $W_{1a}\ll
W_2\ll W_{1b}$, the power law regimes would be as follows: (i) a power
law regime corresponding to the angle $\theta$ between the segments
$W_{1a}$ and $W_2$, (ii) a power law regime corresponding to the angle
$\pi-\theta$ between segments $W_2$ and $W_{1b}$, (iii) a point-line
power law regime entered when $\ell\gg W_{1b}$, and (iv) the 2-points
regime. Figure~\ref{pOverlap} is a plot of $P(\ell|r)$ which shows this behavior.

\subsection{Non-Coplanar Lines}

For non-coplanar lines, for $l\gg r$, the fact that the lines are not
co-planar should be irrelevant; what is relevant is the effective angle
between the lines. This angle is obtained by sliding the lines toward
each other along the line of shortest Euclidean distance between the
lines (without changing their orientations) until they touch; the lines
are then coplanar and the angle between them determines the behavior of
$P(\ell|r)$. Figure~\ref{pNonCoPlanar} contains plots for two
configurations which illustrate this: (i) two coplanar lines with $r=1$,
$\theta=90^\circ$, and $W=256$, and (ii) the same configuration with the
second line translated out of the plane by distance 8.  We see that
while there is some difference in the plots for small $l$, the slope of
the 2-lines regime is the same for the two plots.

\subsection{Scaling of the Crossover Between Power Law Regimes}

We define the value of $\ell \cong \hat\ell$ at which $P(\ell|r)$
crosses over from one power-law regime to another power-law regime as
the value of $\ell$ where straight lines fit to the power law regimes
,between which the crossover takes place, cross. In Eq.(\ref{e.1}) the
values of $\ell$ at which the lower and upper cutoffs occur scale
independently as $r^{d_{\mbox{\scriptsize min}}}$ and
$L^{d_{\mbox{\scriptsize min}}}$, respectively.  By extension, we would
expect that all characteristic values of the distribution, including
crossovers between different power-law regimes, would also scale as
$X^{d_{\mbox{\scriptsize min}}}$ where X is the length in the system which controls the
crossover. Thus, in analogy with the scaling of the most probable value
of $\ell, \ell^\ast$,
\begin{equation}
\label{e.6}
l^\ast=cr^{d_{\mbox{\scriptsize min}}},
\end{equation}
we would, in fact, expect that the value of $\ell$, $\hat\ell$ at which
$P(\ell|r)$ crosses over from 2-lines behavior to 2-points behavior
scales as 
\begin{equation}
\label{e.7a}
\hat\ell=c_1(\theta)r_{\mbox{\scriptsize max}}^{d_{\mbox{\scriptsize
min}}}, 
\end{equation}
where
\begin{equation}
\label{e.7x}
r_{\mbox{\scriptsize max}}=r+2W\sin(\theta/2)
\end{equation}
is the maximum Euclidean distance between the two lines and
$c_1(\theta)$ is a slowly varying function of $\theta$.  In order for a 2-lines
regime to exist, the 2-lines regime cutoff $\hat\ell$ must be greater
than $\ell^\ast$, the maximum value of the distribution. That is,
\begin{equation}
\label{e.15}
c_1[r+2W\sin(\theta/2)]^{d_{\mbox{\scriptsize
min}}}>cr^{d_{\mbox{\scriptsize min}}},
\end{equation}
which implies
\begin{equation}
\label{e.16}
W>{(c/ c_1)^{1/d_{\mbox{\scriptsize min}}}-r\over{2\sin(\theta/2)}}.
\end{equation}

In Figs.~\ref{pEqCutoff}(a), (b), and (c), the insets contain plots of
$\hat\ell$ vs.~$r_{\mbox{\scriptsize max}}$.  For $\theta=3^\circ$, the
scaling exponent is consistent with Eq.(\ref{e.7a}) but for $\theta=29^\circ$
and 180 the scaling exponent is $1.0\pm 0.1$.

\begin{figure}
\centerline{
\xsize
\epsfclipon
\epsfbox{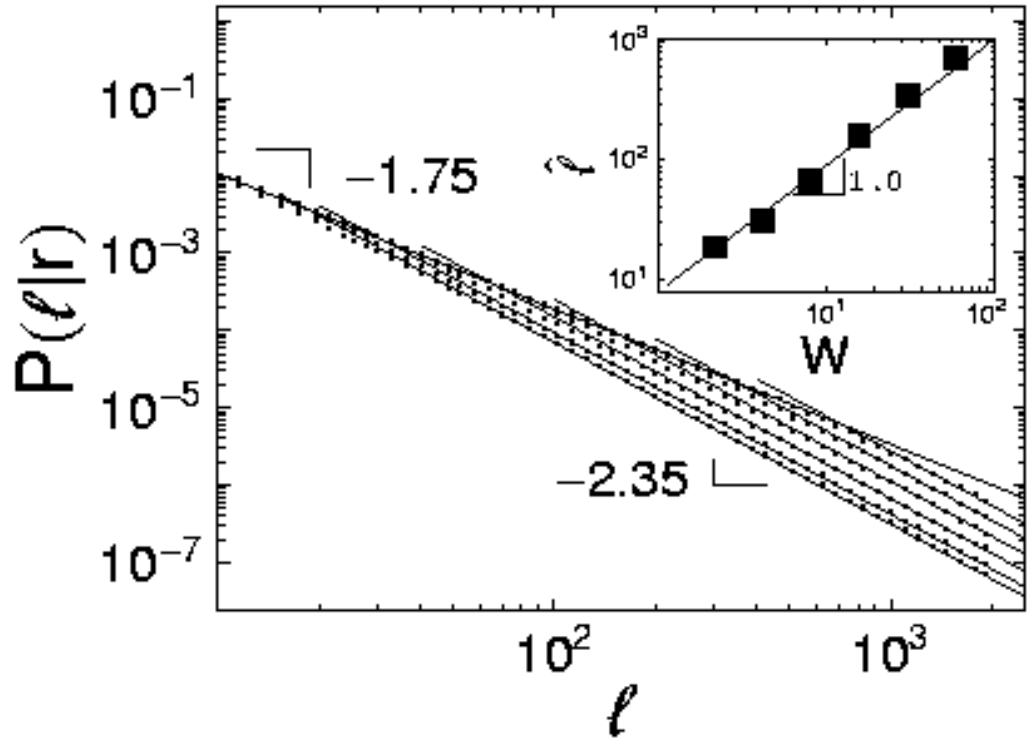}
}
\caption{$P(\ell | r)$ vs $\ell$ for configuration of a point and a line
in which the point is closest to the end of the line with $r=1$ and $W=$
(from bottom to top) 2, 4, 8, 16, 32, 64, and 128. For all plots, the
larger the value of $W$, the larger the value of $\ell\protect\cong\hat\ell$
at which the behavior changes from point-line behavior to 2-points
behavior. The inset plots $\hat\ell$ vs.~$W$.}
\label{pPtLine2}
\end{figure}

Using the same reasoning which led to Eq. (\ref {e.7a}), we would expect
the crossover from point-line to 2-points behavior to scale as

\begin{equation}
\label{e.7b}
\hat\ell=c_2 W^{d_{\mbox{\scriptsize
min}}}, 
\end{equation}
because $W$ is the length which controls this crossover; as seen in
 Fig.~\ref{pPtLine2}, the larger the value of $W$, the larger the value
 of $\ell$ at which the crossover from point-line to 2-points behavior
 occurs.  However, as seen in the inset in Fig.~\ref{pPtLine2}, the
 crossover length scales with an exponent $1.0\pm 0.1$ not
 $d_{\mbox{\scriptsize min}}$.

\begin{figure}

\centerline{
\xsize
\epsfclipon
\epsfbox{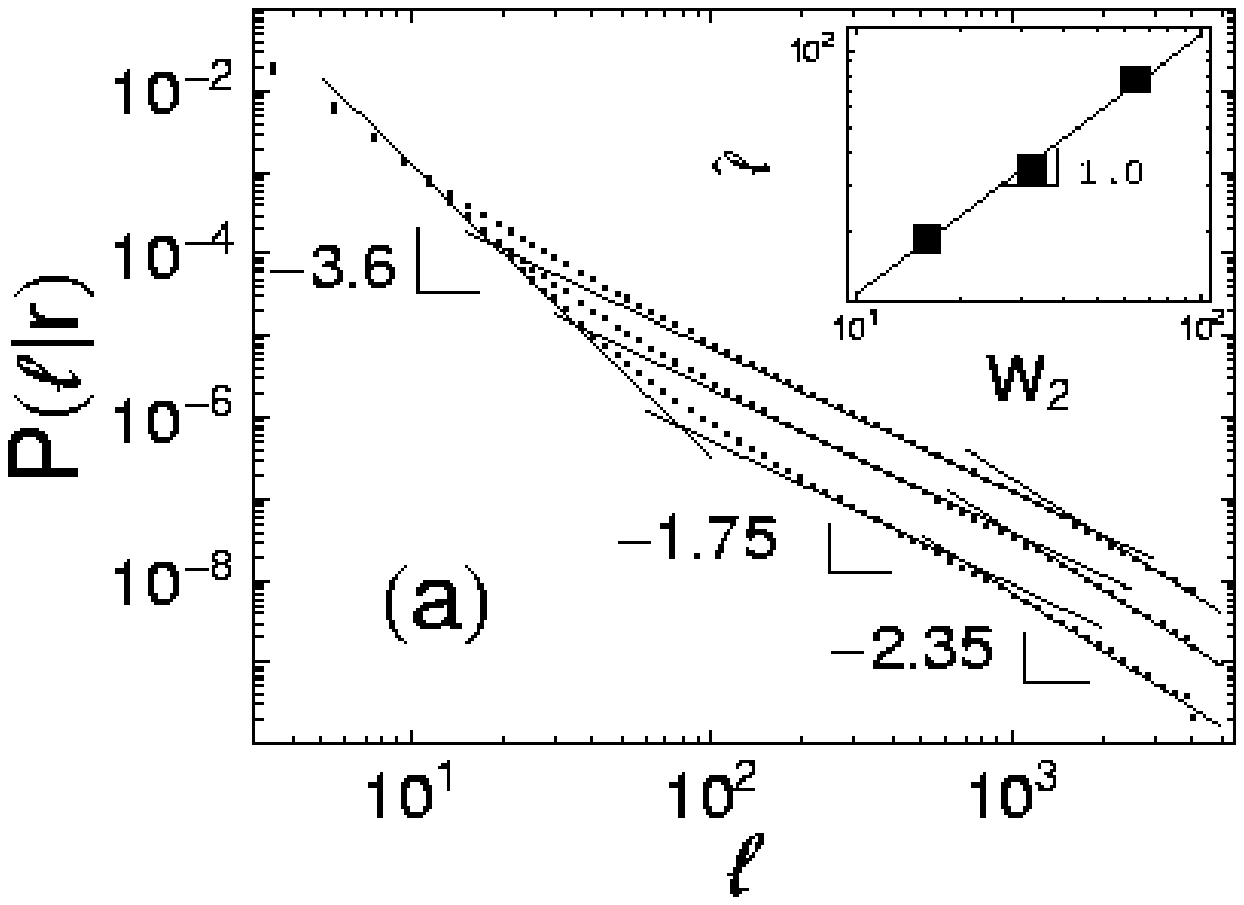}
}

\centerline{
\xsize
\epsfclipon
%\epsfbox{pDiff28.eps}
}

\centerline{
\xsize
\epsfclipon
\epsfbox{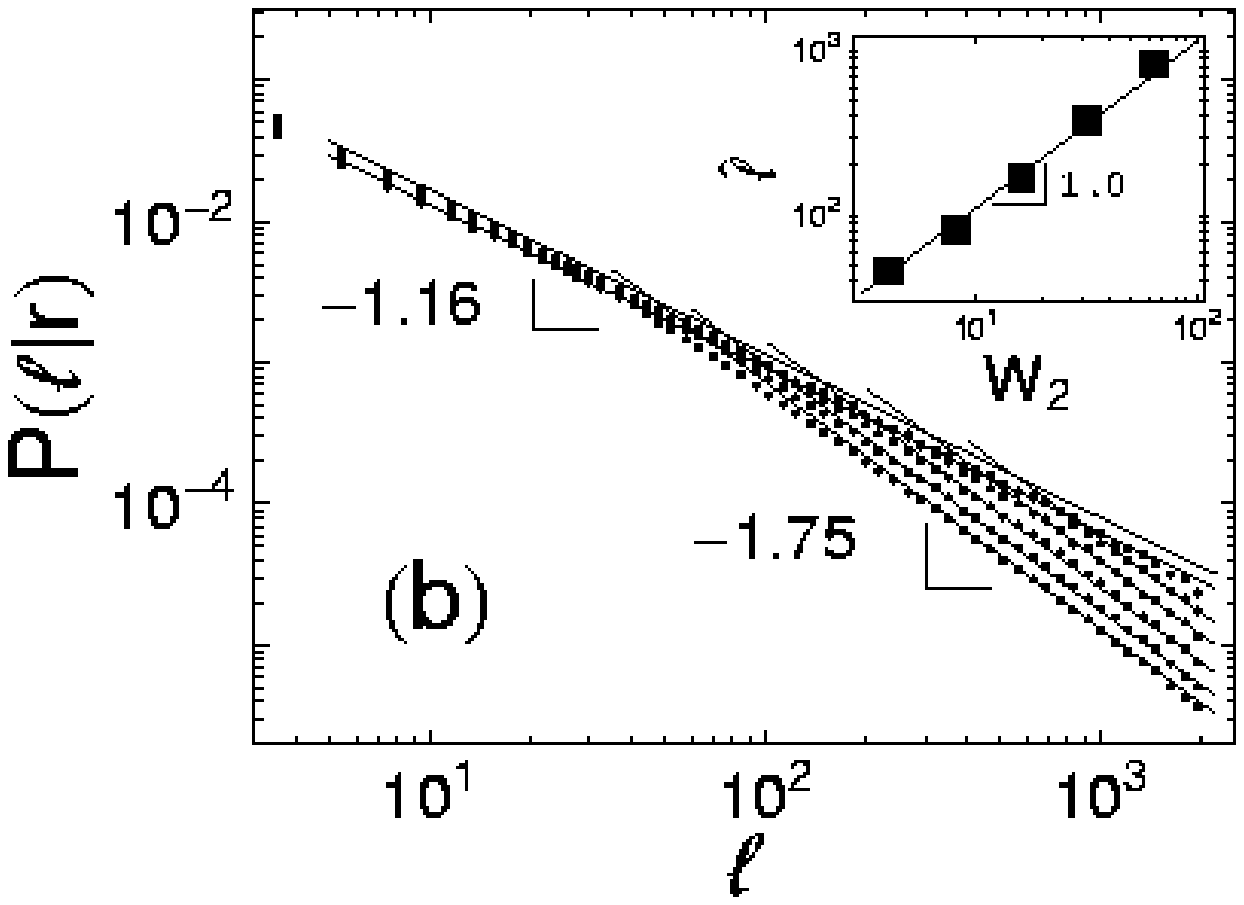}
}
\caption{$P(\ell | r)$ vs $\ell$ for configurations of two lines of
different lengths with $r=1$ and $W_1=128$. (a) $\theta=7^\circ$, $W_2=$ (from
top to bottom) 16, 32, and 64. Three power law regimes can be seen: the
2-lines regime, the point-line regime and the 2 points regime. (b)
$\theta=180^\circ$, $W_2=$ (from bottom to top) 4, 8, 16, 32, 64, and
128. Only the first two power law regimes can be seen: the 2-lines
regime and the point-line regime (the 2-points regime would require even
larger values of $\ell$).}
\label{pDiff}
\end{figure}

\begin{figure}

\centerline{
\xsize
\epsfclipon
\epsfbox{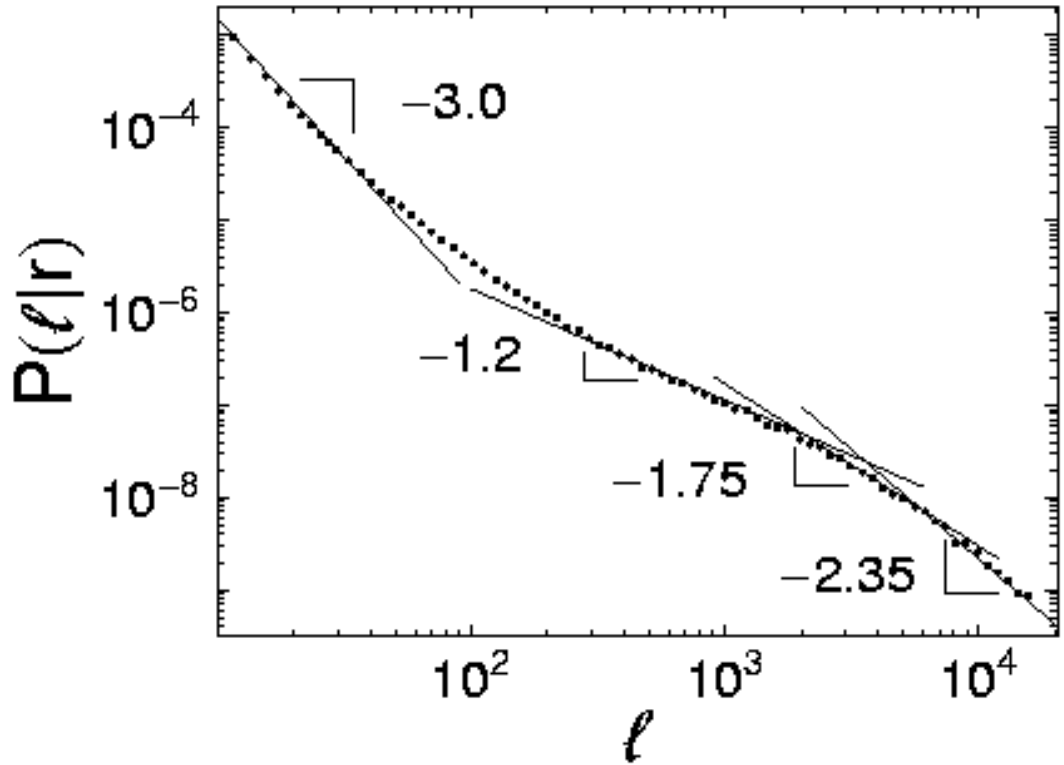}
}
\caption{$P(\ell | r)$ vs $\ell$ for configurations of two lines of
different lengths which ``overlap'' [see Fig.~\protect\ref{pPP1}(c)]
with $\theta=7^\circ$, $r=1$, $W_{1a}=32$, $W_{1b}=128$, and
$W_{2}=256$. Four power law regimes are present: the 2-line
($\theta=7^\circ$) regime (slope $-3.0$), the 2-line
($\theta=180^\circ-7^\circ$) regime (slope $-1.2$), the point-line
regime (slope $-1.75$), and the 2-points regime (slope $-2.35$).}
\label{pOverlap}
\end{figure}

\begin{figure}

\centerline{
\xsize
\epsfclipon
\epsfbox{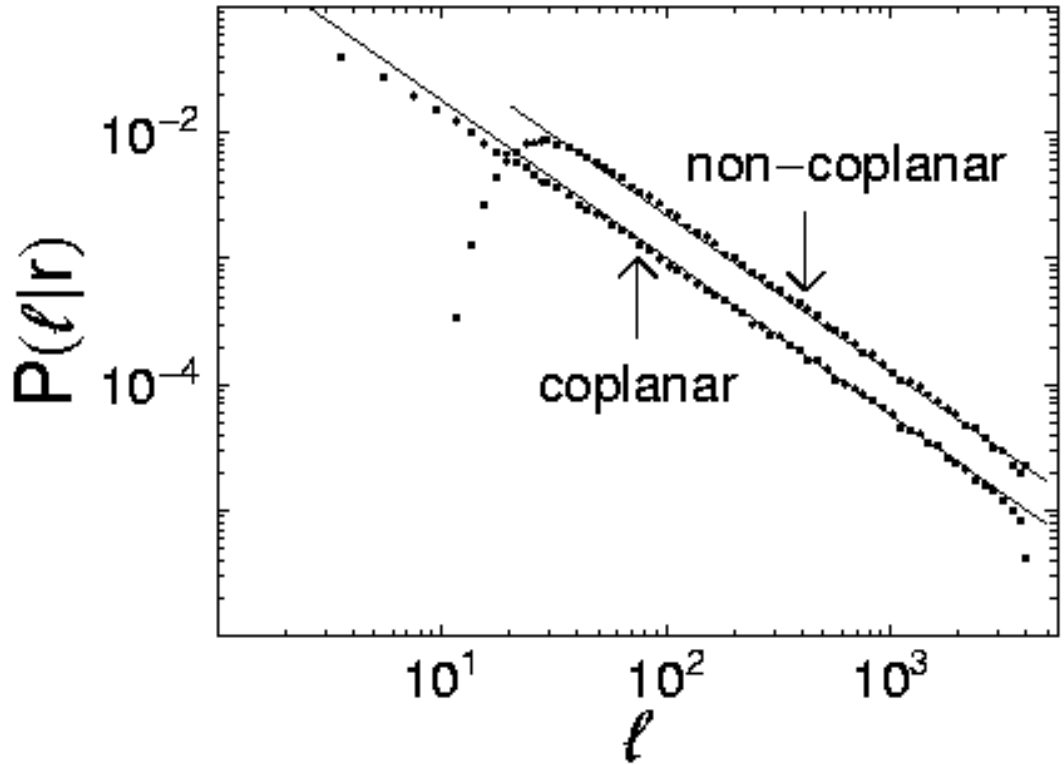}
}
\caption{$P(\ell | r)$ vs $\ell$ for configurations of two lines of
equal length. The co-planar configuration has $r=1$, $\theta=90^\circ$, and
$W=256$ and the lines are co-planar. The non-coplanar configuration is
obtained from the co-planar configuration by moving one of the
lines a distance 8 perpendicular to the plane defined by the coplanar
lines. One sees that for large $\ell$, the power law regimes of the two
plots have the same exponent.}
\label{pNonCoPlanar}
\end{figure}

Finally, we would expect that for different length lines, the
crossover from 2-lines behavior to 2-point behavior would scale as 
\begin{equation}
\label{e.7c}
\hat\ell=c_3(\theta)W_{2}^{d_{\mbox{\scriptsize
min}}}, ~~~~~~~~~~(W_2<W_1)
\end{equation}
because $W_2$ is the length which controls this crossover; as seen in
Fig.~\ref{pDiff}, the larger the value of $W_2$, the larger the value of
$\ell$ at which the crossover occurs.  Again, the insets in
Fig.~\ref{pDiff} indicate that the crossover scales with exponent $1.0\pm 0.1$.

We cannot explain why sometimes the crossover scales with
$d_{\mbox{\scriptsize min}}$ and sometimes it scales with an exponent
about 1.  It is, of course, possible that corrections-to-scaling are
strong and that we are not seeing the true asymptotic behavior of the
scaling of the crossover.  If this is the case, the question still
remains as to why the corrections to scaling are strong in some
configurations and not in others.  This area is a subject for further
study.

\section{Parallel Wells}

\subsection{Simple Configurations}

As with non-parallel wells we first consider the simple configurations
shown in  Fig.~\ref{pPConfig}(a) in which the parallel wells are of the same
length. Figure~\ref{pPar}(a) plots $P(\ell | r)$ vs $\ell$ for $r=1$ and various
$W$.  The initial decay of the plots increases with increasing W because
the longer the wells, the lower the probability for long shortest paths.
Eventually, all plots cross over to a power-law regime with slope
consistent with that for two points.  To see if this initial decay is a
lattice effect, Fig.~\ref{pPar32} plots of the scaled distributions
$r^{d_{\mbox{\scriptsize min}}}P(\ell/r^{d_{\mbox{\scriptsize min}}}|W)$
for various $r$ and $W$ where the aspect ratio,
\begin{equation}
\label{e.10}
R={W\over r},
\end{equation}
is fixed at $R=32$. Changing $r$ and $W$ but keeping $R$ fixed results
in scaling all lengths in the geometry by the same factor and the plots
collapse as expected. Again, we note the initial strong decay of the
distribution followed by a 2-point power-law regime.  The good collapse
for small $x=1/r^{d_{min}}$ indicates that the strong initial decay is
not a lattice effect.  

Because of the small values of $\ell$ at which the crossover
to the 2-point regime occurs it is difficult to differentiate between a
power-law and (stretched) exponential decay. We will proceed as if the
decay were either a power law with slope $\bar g(R)$ or equivalently an
exponential with ``effective slope'' $\bar g(R)$.

One might argue as follows that the initial decay for power parallel
lines must be exponential: Since the 2-lines regime of the probability
distribution for a parallel well configuration must always decay faster
than the 2-lines regime of a configuration with small but non-zero
$\theta$ and since we believe $g_{\ell}(\theta)$ goes to infinity as
$\theta$ goes to zero, the decay for parallel lines must be exponential
(i.e., faster than any power law decay). This, however, need not be the
case. In order for a 2-lines regime to exist, Eq.~(\ref{e.16}) must
hold. So as we decrease $\theta$, we must increase $W$, increasing the
aspect ratio $R$, to maintain a 2-lines regime. But since the effective
slope for parallel wells, $\bar g(R)$ increases with increasing $R$, the
decay can be a power law and still always have a greater slope than the
configuration with small but non-zero $\theta$.

\subsection{Complex Configurations}

\begin{figure}

\centerline{
\epsfxsize=7.0cm
\epsfclipon
\epsfbox{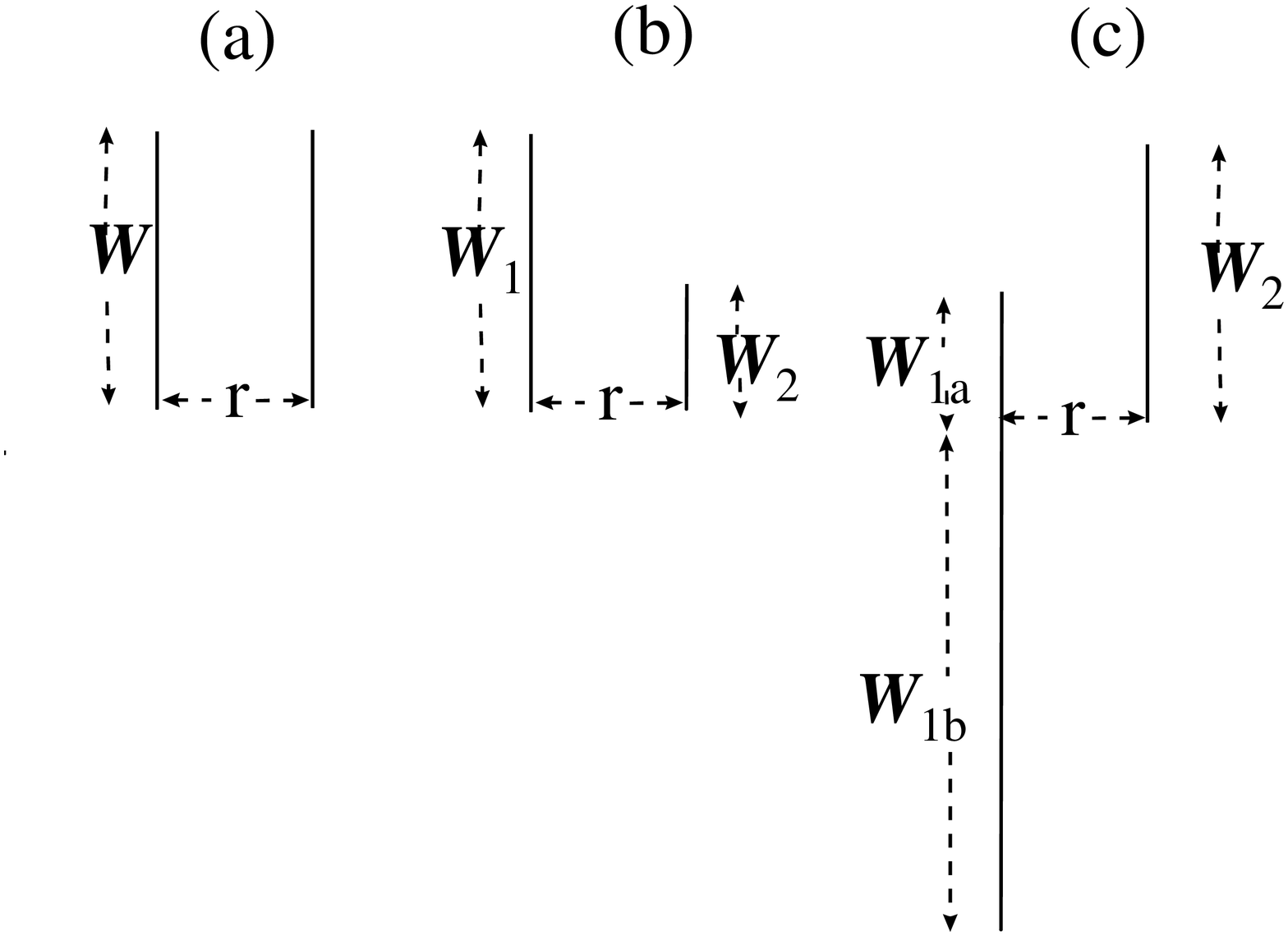}
}

\caption{Example configurations for parallel wells. (a) Simple
configuration of wells of equal length. (b) Configuration of wells of
unequal length ($W_1>W_2$). (c) Configuration in which shortest line
between end of one well does not terminate at end of other
well ($W_{a1}<W_2<W_{1b}$).}
\label{pPConfig}
\end{figure}

\begin{figure}

\centerline{
\xsize
\epsfclipon
\epsfbox{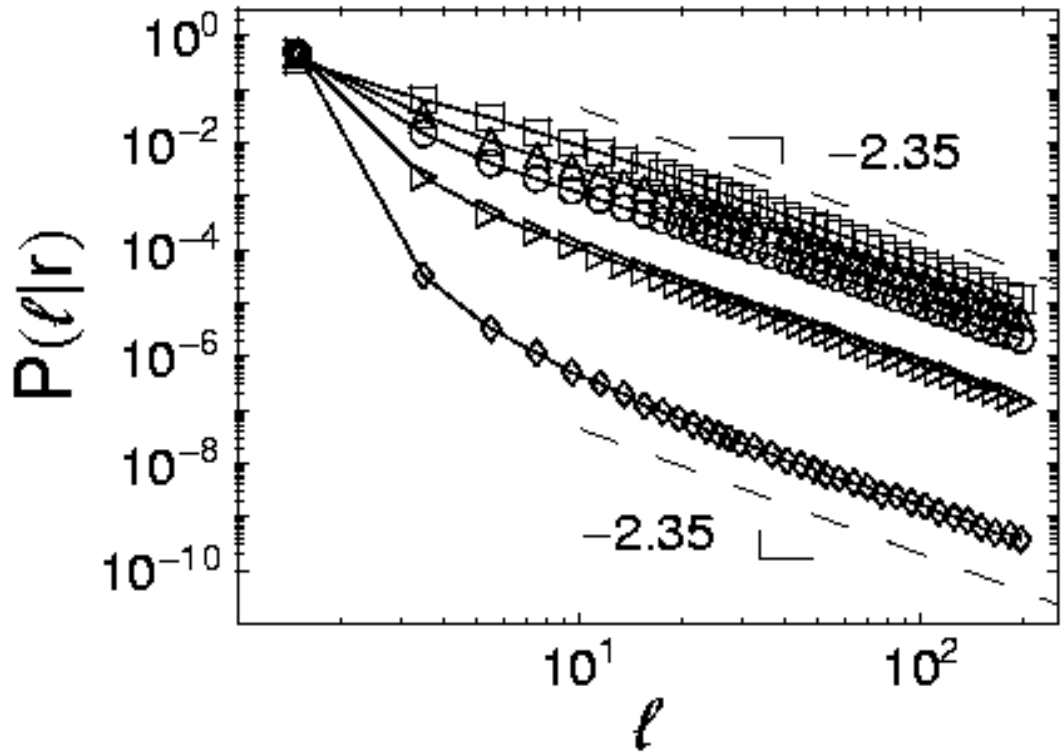}
}
\caption{$P(\ell | r)$ vs $\ell$ for configurations of two parallel
lines of equal length with $r=1$ and $W=$ (from top to bottom) 0 (two
points), 4, 8, 16, and 32. The slopes of the power law regimes of the
plots for all configurations is the same but the initial decay of the
plots increases with increasing $W$.}
\label{pPar}
\end{figure}

\begin{figure}

\centerline{
\xsize
\epsfclipon
\epsfbox{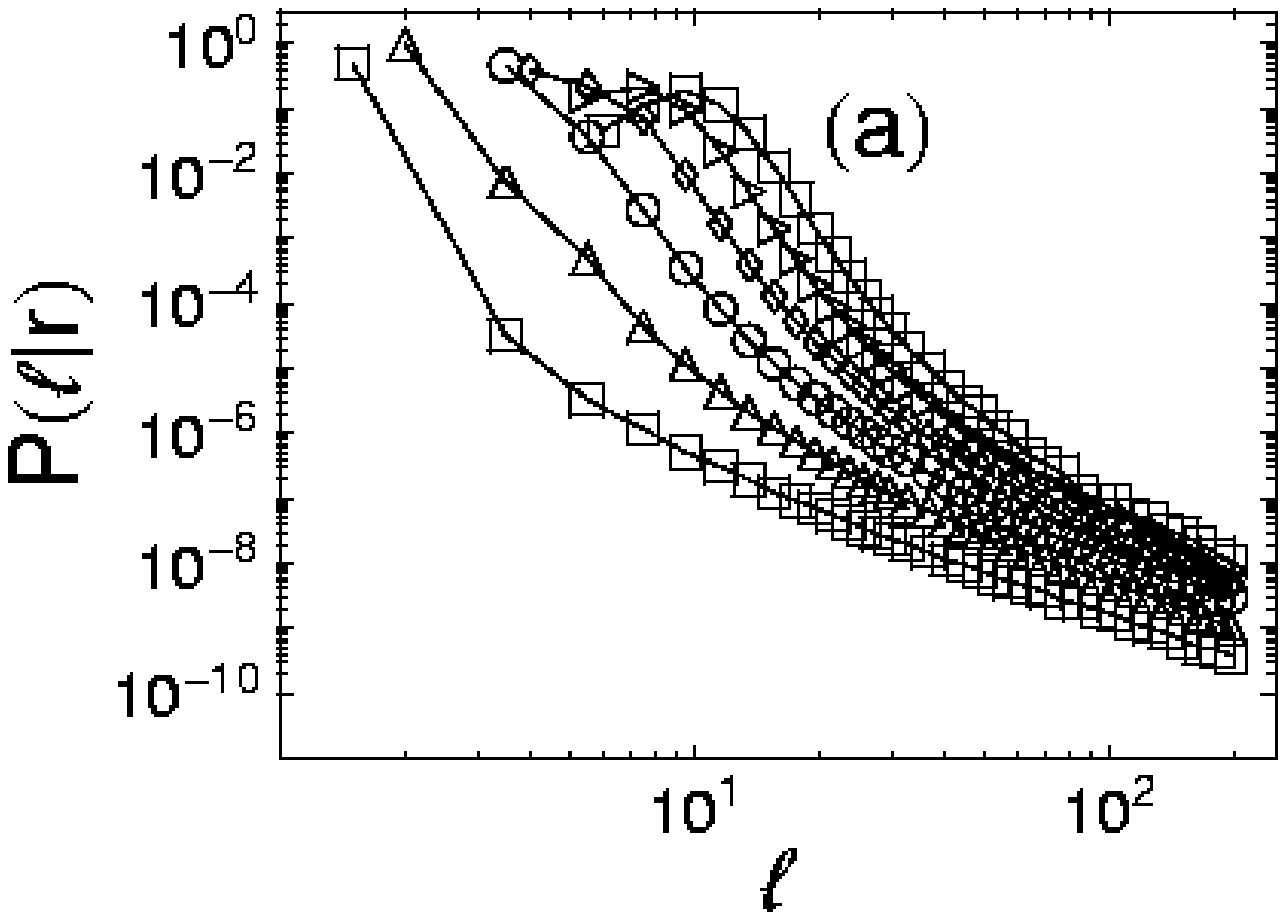}
}

\centerline{ 
\xsize 
\epsfclipon
\epsfbox{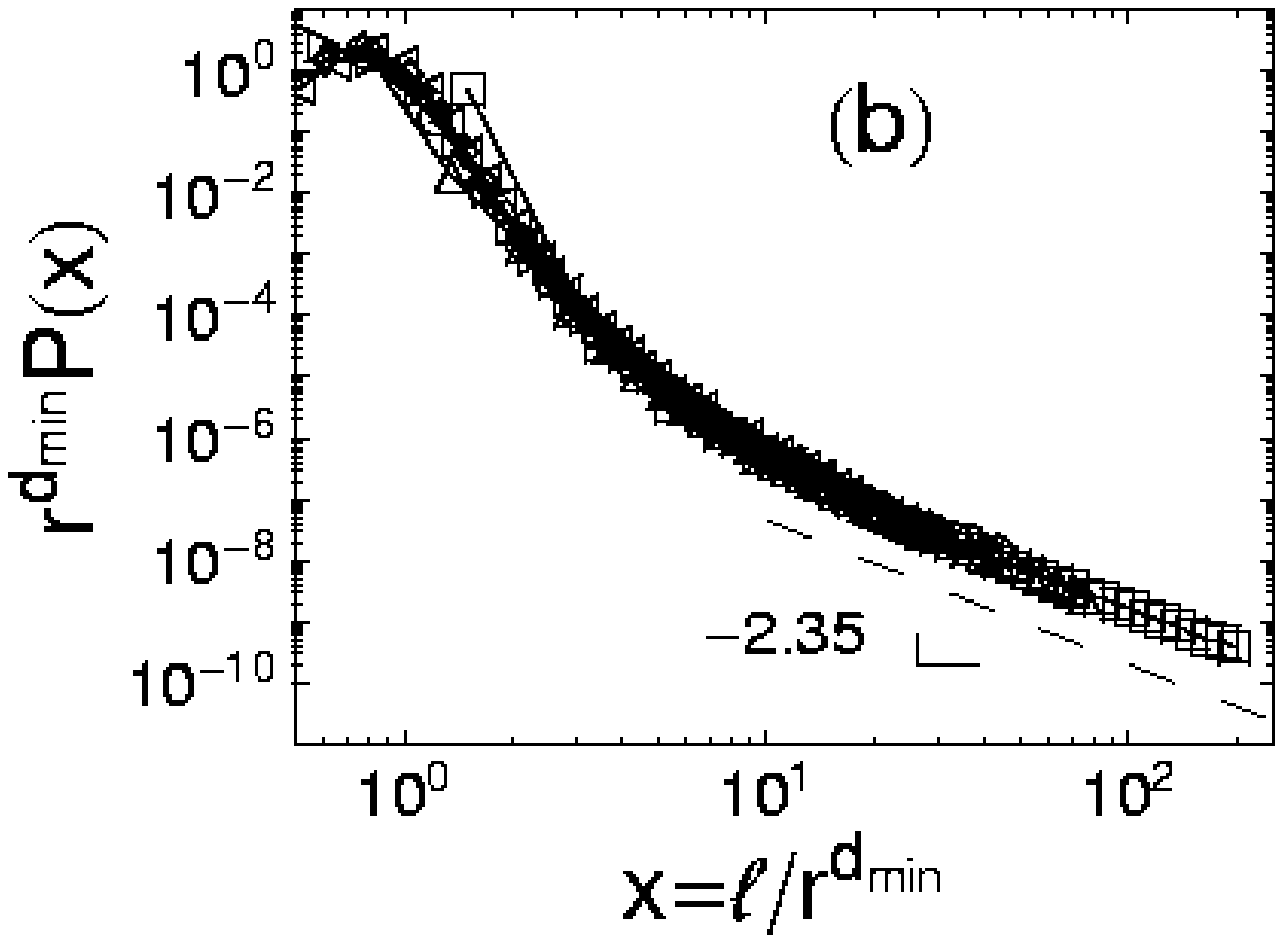} 
}
\caption{$P(\ell | r)$ vs $\ell$ for configurations of two parallel
lines of equal length with $(W,r)=$ (from top to bottom) (32,1), (64,2),
(96,3), (128,4), (160,5), and (196,6) (b) plots of (a) scaled with the
variable $x=\ell/r^{d_{\mbox{\scriptsize min}}}$. The plots in (b)
collapse nicely as would be expected since they all have the same
aspect ratio, $W/r$. The good collapse for small $x$ indicates that
the small $x$ behavior
is not a lattice effect.}  
\label{pPar32}
\end{figure}

The treatment of the more complex configurations shown in Figs.~\ref{pPConfig}(b) and
(c) follows that of non-parallel wells. $P(\ell)$ for configurations of
the type in Fig.~\ref{pPConfig}(b) would contain an initial 2-line regime with slope
$\bar g(R=W_2/r)$, a point-line regime, and finally a two-point
regime. $P(\ell)$ for configurations of the type in Fig.~\ref{pPConfig}(c), with
$W_{1b}\ll W_2\ll W_{1a}$ would contain an initial 2-line regime with
slope $\bar g(R=W_{1b}/r)$, a 2-line regime with slope $g_{\ell}(\theta=\pi)$, a point-line regime, and a 2-point regime.

\subsection{Quasi-Euclidean Regime}

\begin{figure}

\centerline{
\xsize
\epsfclipon
\epsfbox{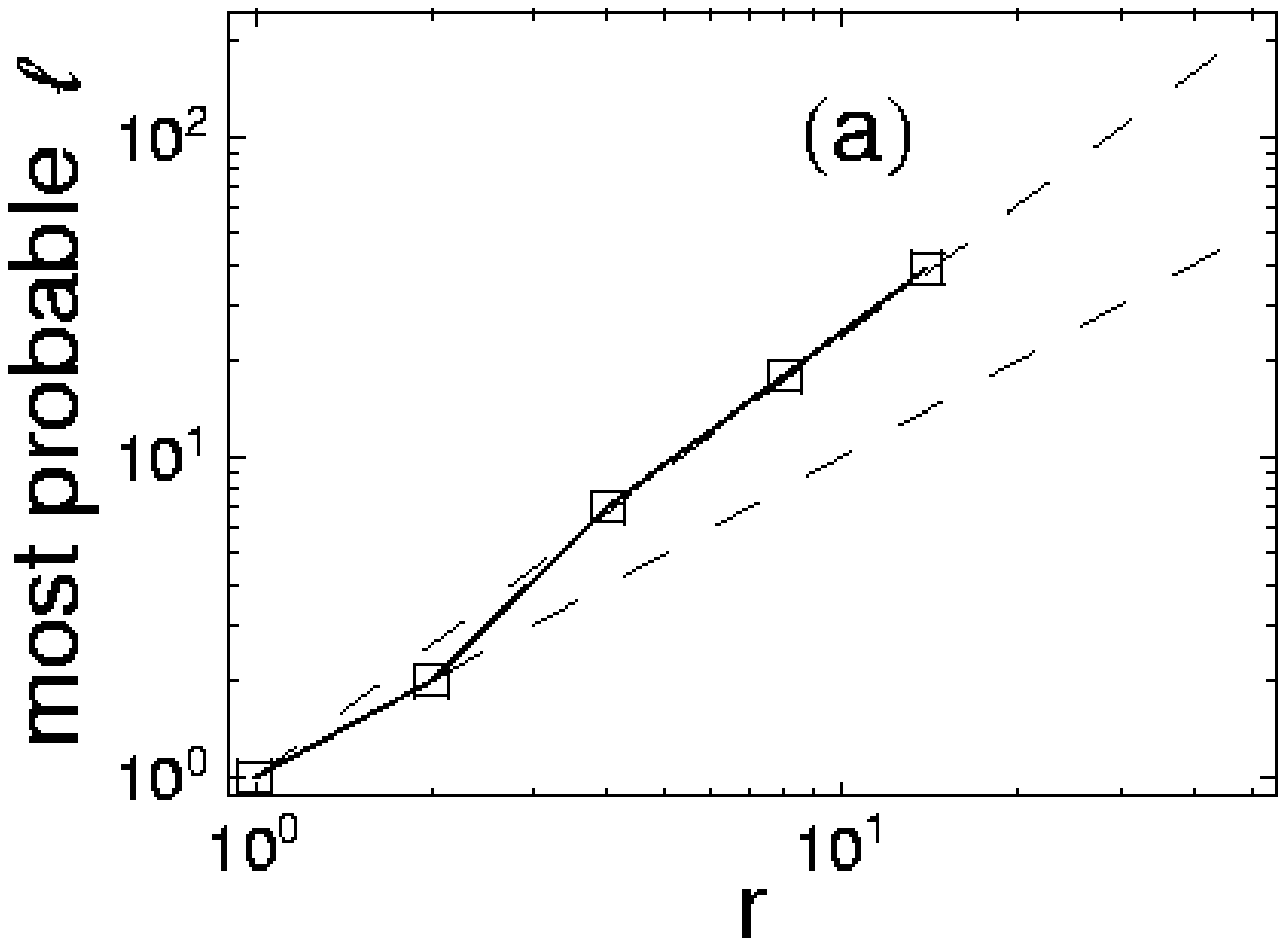}
}

\centerline{
\xsize
\epsfclipon
\epsfbox{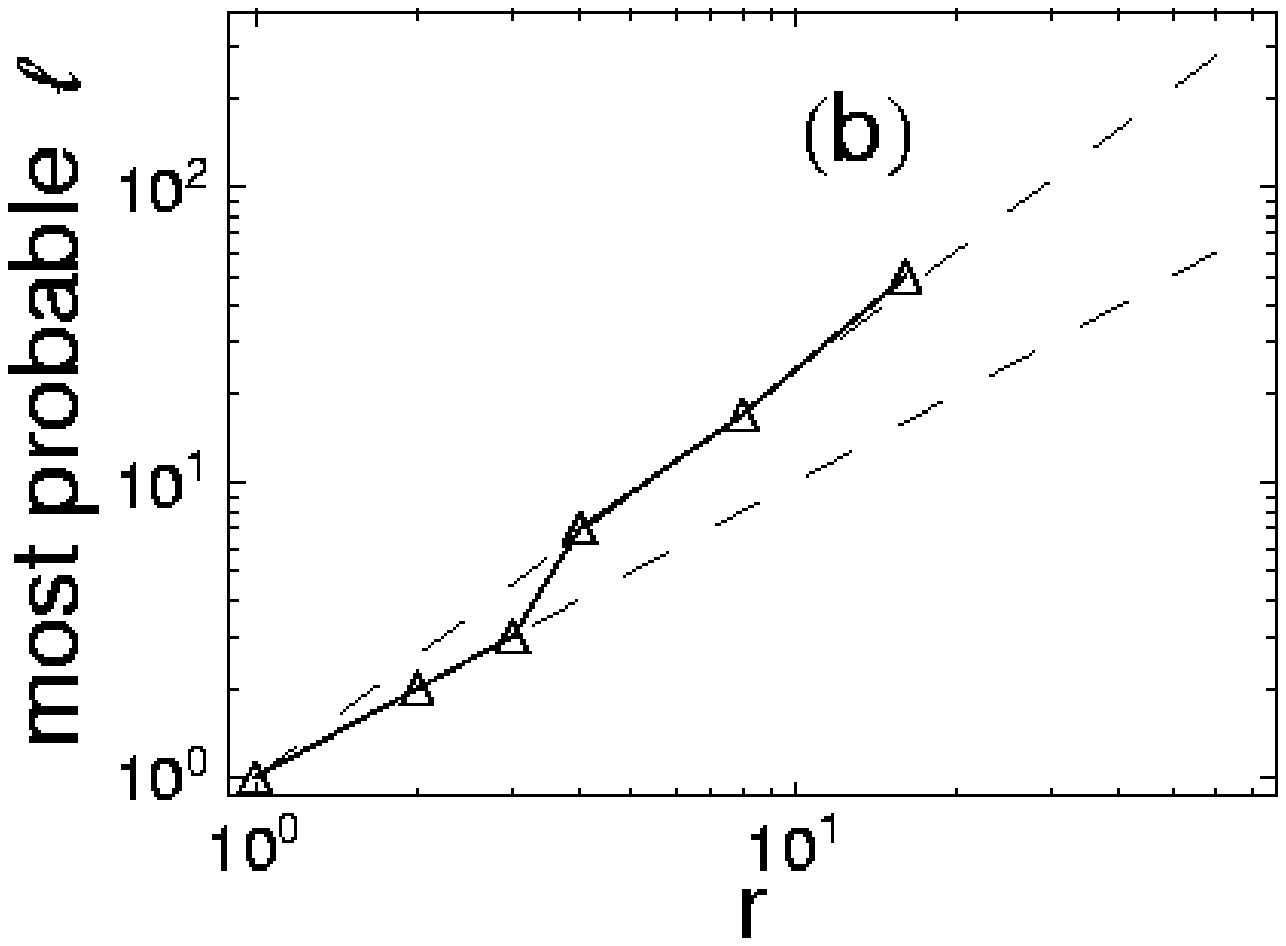}
}

\centerline{
\xsize
\epsfclipon
\epsfbox{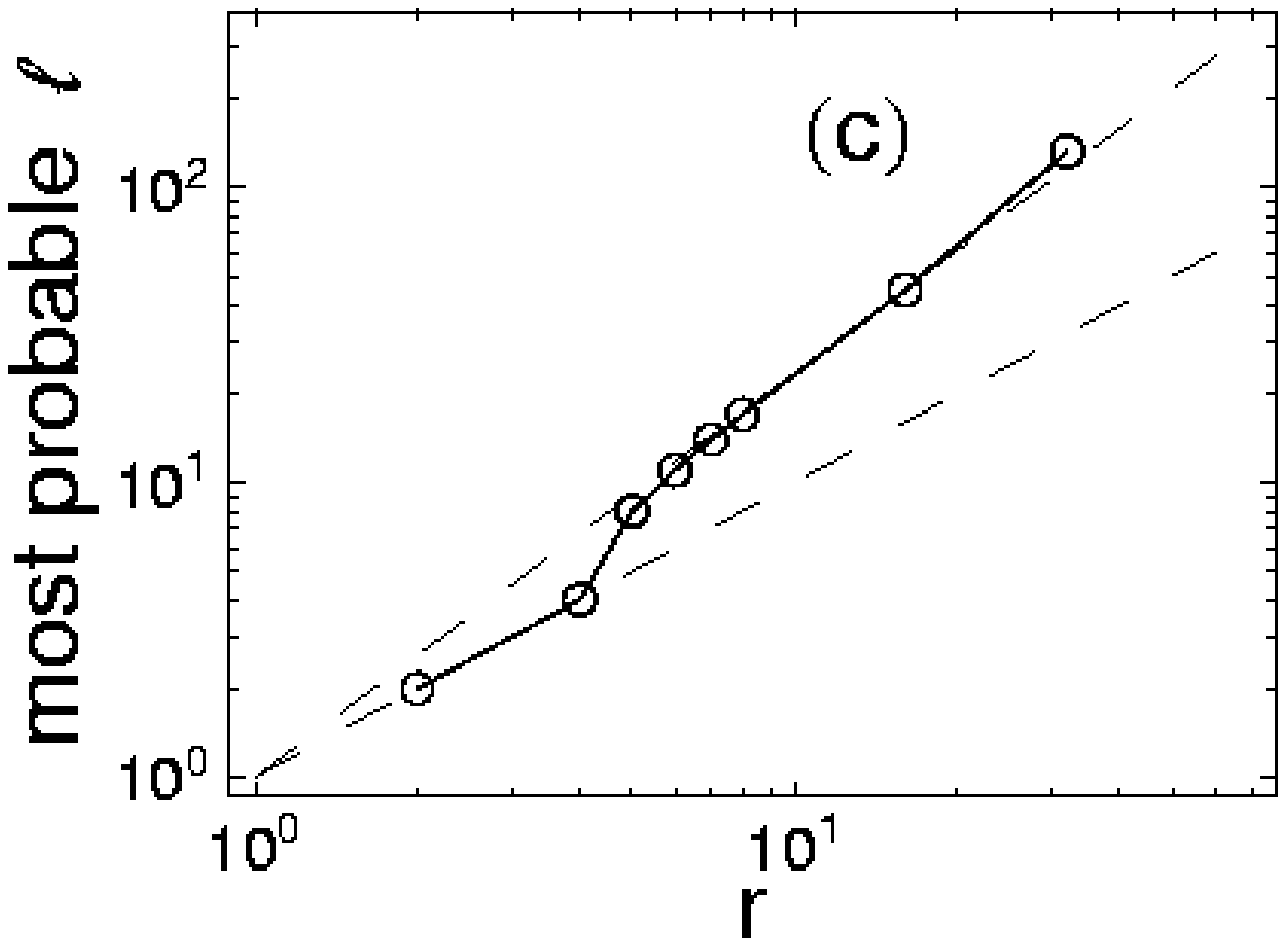}
}

\centerline{
\xsize
\epsfclipon
\epsfbox{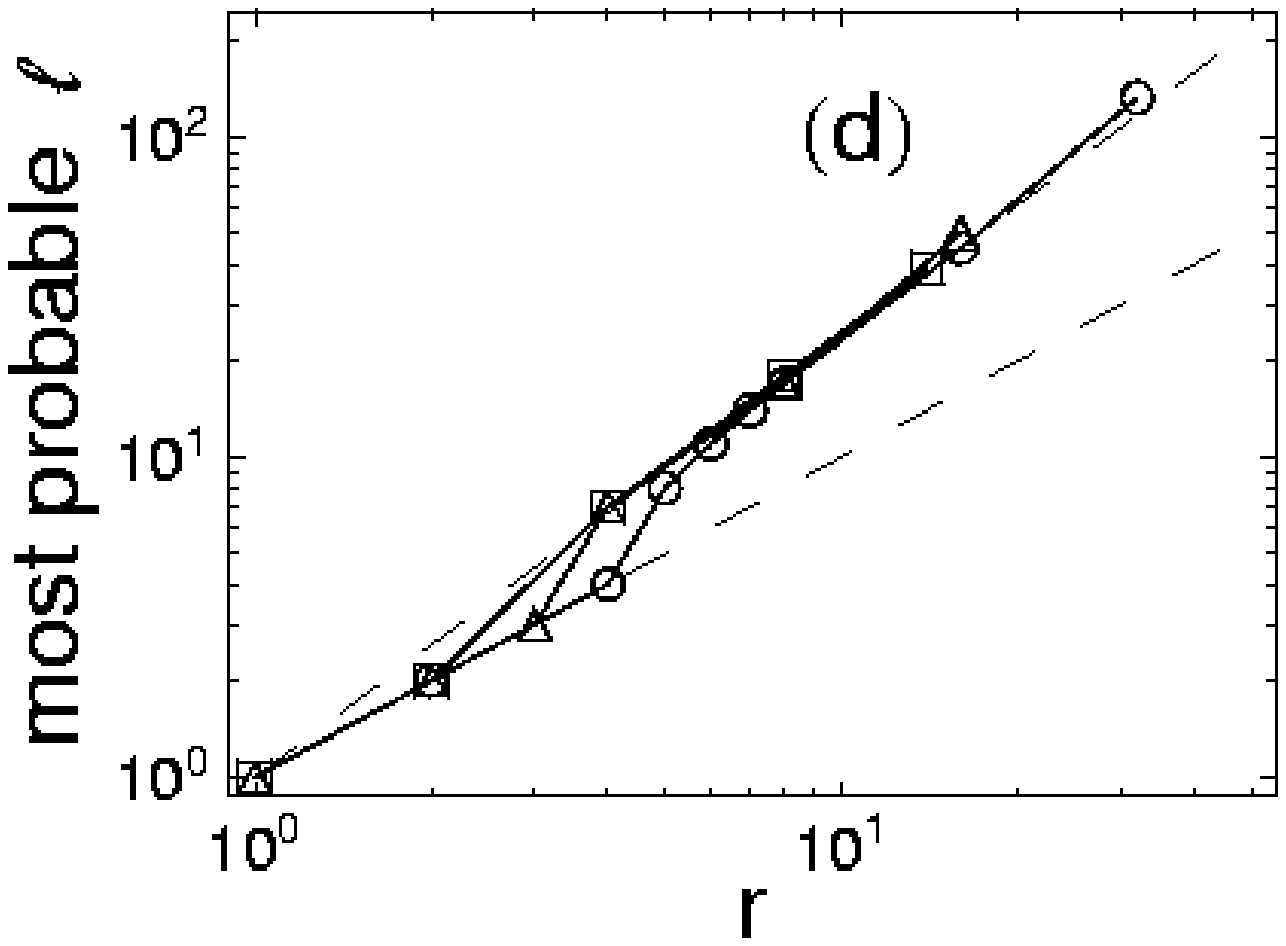}
}
\caption{Most probable $\ell$ vs $r$ for configurations of two parallel lines
of equal length. (a) $W=16$, $r=1$, 2, 4, 8, and 14. (b) $W=32$, $r=1$,
2, 3, 4, 8, and 16. (c) $W=64$, $r=2$, 4, 5, 6, 7, 8, 16, and 32. (d)
Combined plot of (a), (b) and (c). The upper and lower dashed lines have
slope $d_{\mbox{\scriptsize min}}$ (1.374) and 1.0, respectively. The
larger the value of $W$, the larger the value of $r$ at which scaling
crosses over from Euclidean behavior to fractal behavior.}
\label{pMax32}
\end{figure}

\begin{figure}

\centerline{
\xsize
\epsfclipon
\epsfbox{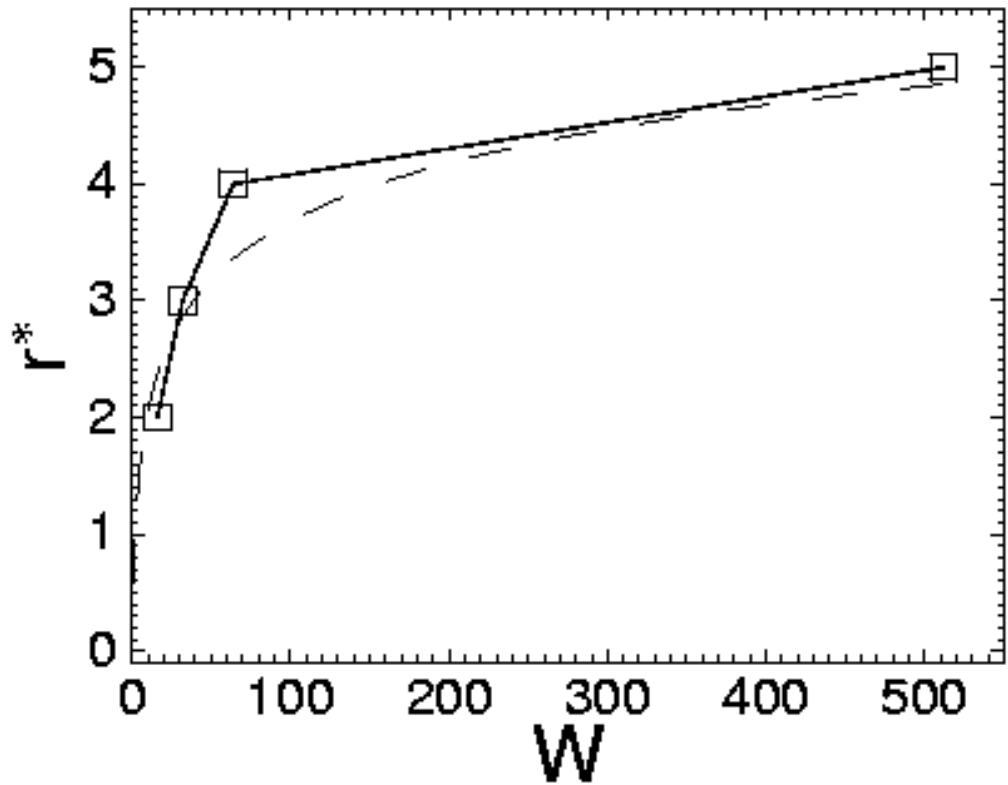}
}
\caption{Value of $r$ at which behavior changes from Euclidean to
fractal, $r^\ast$, $W$. The dashed line is a prediction of
Eq.~(\protect\ref{e.51}).}
\label{pMax32t}

\end{figure}

When the length of the wells is very large and the distance between the
wells is small the behavior of the most probable shortest path between
the wells is the same as in a Euclidean space where p=1 and all bonds
are occupied. This can be seen in Fig.~\ref{pMax32} in which we plot
$\ell^\ast$, the most probable value of the shortest path, versus $r$
for various lengths $W$. For long enough wells, there is a regime of $r$
in which
\begin{equation}
\label{e.12x}
\ell^\ast=r,
\end{equation}
as one would expect in Euclidean space in which the shortest path is a
straight line path of occupied bonds. As also seen in Fig~\ref{pMax32}., for a
given well length, as $r$ increases, there is a value of $r$, $r^\ast$,
at which the behavior crosses over to that of 3D percolation. We can
develop a simple expression for $r^\ast$ as follows: The probability
that all bonds in a straight line path between two wells separated by
distance $r$ are occupied is $p_c^r$. The probability that one or more
bonds in the straight line path is not occupied is thus $1-p_c^r$ and
the probability that one or more bonds in the $W$ straight line paths
between the wells are not occupied is $(1-p_c^r)^W$. The probability
that at least one straight line path has all bonds occupied is then
\begin{equation}
\label{e.50}
P(r,W)=1-(1-p_c^r)^W.
\end{equation}
The shortest path will exhibit Euclidean behavior, i.e., $\ell^\ast=r$
when $P(r,W)$ is of the order unity. Setting $P(r^\ast,W)=a$ in
Eq.~(\ref{e.50}), we find
\begin{equation}
\label{e.51}
r^\ast={\ln(1-a^{1/W})\over\ln p_c}.
\end{equation}
In Fig.~\ref{pMax32t} we plot the observed values of $r^\ast$ and the values
predicted by Eq.~(\ref{e.51}) with a value of $a=0.55$ which gives the
best fit to the observed values.

\section{Relationship between Parallel Wells and ``Close to Parallel''
Wells}

For a given $r$, we expect that a configuration with small but non-zero
angle will have a  power-law regime slope very close to the
(effective) power-law regime slope of a configuration of parallel lines
with the same $W$. This at first leads to a seeming paradox: if we
increase or decrease $W$, but keep the angle of the non-parallel wells
fixed, the slope of the 2-lines regime of the non-parallel well
configuration doesn't change as discussed in Section III.A.1.  However,
if we consider this configuration as a parallel configuration, changing
W changes the aspect ratio which changes the power-law regime slope as
discussed in Section IV.A.  This seeming inconsistency is resolved as
follows: on the one hand, for a 2-lines regime to exist, W must be at
least as large as the value given by Eq.~(\ref{e.16}.  If $W$ is too
small, there will be no 2-lines regime and both the parallel and small
angle configurations will look like the configuration for 2 points.  On
the other hand, if $W$ is increased, keeping $r$ and $\theta$ fixed, the
greater the deviation from parallel lines and there is no reason why the
parallel and small-angle configurations should have the same slopes in
their power-law regimes.

Thus, only for the very small range of W for which the power-law regime
exists and for which the configuration with small but non-zero $\theta$
is ``close to parallel''(i.e. the difference between the values of $r$
and $r_{max}$ is small) should the slopes of the parallel configuration
and the configuration with small but non-zero $\theta$ be equal.  That
is,
\begin{equation}
\label{e.18}
\bar g[W/r]\approx g_{\ell}(\theta),
\end{equation}
where $\bar g(R)$ is defined in Section IV.A.

\section{Discussion and Summary}

Motivated by the need to more realistically model the geometries found
in oil recovery activities, we have determined the scaling form for the
distribution of shortest paths between two lines in 3 dimensional
percolation systems.  Using simple scaling arguments we explained the
rich fractal behavior of the shortest path in these systems.  A number
of open questions, however, remain:

\begin{itemize}

\item[{(i)}] From first principles, can one develop an expression for
$g_{\ell}(\theta)$?  An exact expression for $g_{\ell}$ for two points in
2-dimensions was obtained by Ziff\cite{Ziff99} using conformal
invariance arguments.  Possibly this approach could be extended to find
$g_{\ell}$ for point-line and 2-line configurations, at least in
2-dimensions.

\item[{(ii)}] How is the fact that the crossover from one power-law
regime to another does not scale with the exponent $d_{\mbox{\scriptsize
min}}$ explained?  Is this simply an artifact of corrections-to-scaling
which would disappear if we could simulate much larger systems or is the
scaling of the crossover actually anomalous in certain configurations?

\end{itemize}

%\subsubsection*{Acknowledgements}
\section*{Acknowledgements}

We thank Sergey Buldyrev, Nikolay Dokholyan, Youngki Lee, Eduardo Lopez,
Peter King, and Luciano da Silva for helpful discussions and BP-Amoco
for financial support.

%\end{thebibliography}

\end{multicols}  

\newpage

%%%%%%%%%%%%%%%%   configuration(overview)

%%%%%%%%%%%%%%%%%%

%%%%%%%%%%%%%%%%%   equal length - different slopes
%%%%%%%%%%%%%%%%%%

\newpage
%%%%%%%%%%%%%%%%%   plot of gl(theta) 

%%%%%%%%%%%%%%%%%%

\newpage
%%%%%%%%%%%%%%%%%   equal length - cutoffs 

%%%%%%%%%%%%%%%%%%%%%%%%%%%%%%

\newpage
%%%%%%%%%%%%%%%%   configuration(point line)

%%%%%%%%%%%%%%%%% point-line

%%%%%%%%%%%%%%%%%%%%%
\newpage

%%%%%%%%%%%%%%%%%%%%   different lengths 

%%%%%%%%%%%%%%%%%%%%%%%%%%%%%%%%%%%
\newpage
%%%%%%%%%%%%%%%%   non-coPlanar

%%%%%%%%%%%%%%%%%

%%%%%%%%%%%%%%%%   configuration(parallel lines)

\newpage

%%%%%%%%%%%%%%%%   parallel (different aspect ratios

\newpage

%%%%%%%%%%%%%%%%   parallel(all with aspect ratio=32

\newpage

%%%%%%%%%%%%%%%%   parallel(plots of most probable(max) versus aspect ratio

%%%%%%%%%%%%%%%%   parallel(theory of cross over to Euclidean behavior)

\end{document}